\documentclass[secnumarabic,amssymb,nobibnotes,aps,prb,preprint,letterpaper,preprintnumbers,amsmath,superscriptaddress,floatfix]{revtex4-2}

\usepackage{wrapfig}
\usepackage{amsmath}
\usepackage{enumitem}
\usepackage{multirow}
\usepackage[letterpaper,margin={1in,1in}]{geometry}
\usepackage{graphicx}
\usepackage{xcolor}
\usepackage{epstopdf}
\usepackage{lineno}
\usepackage{xcolor}
\usepackage{hyperref}
\usepackage[normalem]{ulem}
\setcitestyle{super}

\begin{document}

\title{Nonreciprocal directional dichroism at telecom wavelengths} 

\author{K. Park}
\affiliation{Department of Chemistry, University of Tennessee, Knoxville, Tennessee 37996 USA}

\author{M. O. Yokosuk}
\affiliation{Department of Chemistry, University of Tennessee, Knoxville, Tennessee 37996 USA}

\author{M. Goryca}
\affiliation{National High Magnetic Field Laboratory, Los Alamos, New Mexico 87545 USA}

\author{J. J. Yang}
\affiliation{Department of Physics, New Jersey Institute of Technology, Newark, New Jersey 07102 USA}

\author{S. A. Crooker}
\affiliation{National High Magnetic Field Laboratory, Los Alamos, New Mexico 87545 USA}

\author{S. -W. Cheong}
\affiliation{Department of Physics and Astronomy, Rutgers University, Piscataway, New Jersey 08854 USA}
\affiliation{Rutgers Center for Emergent Materials, Rutgers University, Piscataway, New Jersey 08854 USA}

\author{K. Haule}
\affiliation{Department of Physics and Astronomy, Rutgers University, Piscataway, New Jersey 08854 USA}

\author{D. Vanderbilt}
\affiliation{Department of Physics and Astronomy, Rutgers University, Piscataway, New Jersey 08854 USA}

\author{H. -S. Kim}
\email{heungsikim@kangwon.ac.kr}
\affiliation{Department of Physics and Institute of Quantum Convergence Technology, Kangwon National University, Chuncheon 24341 Korea}

\author{J. L. Musfeldt}
\email{musfeldt@utk.edu}
\affiliation{Department of Chemistry, University of Tennessee, Knoxville, Tennessee 37996 USA}
\affiliation{Department of Physics and Astronomy, University of Tennessee, Knoxville, Tennessee 37996 USA}


\begin{abstract}

Magnetoelectrics with ultra-low symmetry and spin-orbit coupling are  well known to display a number of remarkable properties including
nonreciprocal directional dichroism. As a polar and chiral magnet, Ni$_3$TeO$_6$ is predicted to host this effect in three fundamentally different configurations, although only two have been experimentally verified. 
Inspired by the opportunity to unravel the structure-property relations of such a 
unique light-matter interaction, we combined magneto-optical spectroscopy and first-principles calculations to reveal nonreciprocity in the toroidal geometry and compared our findings with  the chiral configurations. We find that formation of Ni toroidal moments is responsible for the largest effects near 1.1 eV   - a  tendency that is captured by our microscopic model and computational implementation.   
At the same time, we demonstrate deterministic control of nonreciprocal directional dichroism in Ni$_3$TeO$_6$ across 
the entire telecom wavelength range.  
This discovery will accelerate the development of photonics applications  that take advantage of unusual symmetry characteristics.

\end{abstract}

\maketitle


\newpage

\section*{Introduction}

The combination of strong spin-orbit coupling and ultra-low symmetry gives rise to many unique properties in materials. 
One of the most curious is
nonreciprocal directional dichroism or `one-way transparency'.\cite{Kezsmarki2014,Tokura2018} 
Spectroscopically, the effect
occurs near a magnetoelectric excitation, often in the THz\cite{Kezsmarki2015,Lee2016,Bordacs2015,Yu2018,Sheu2019,Bordacs2012, Kezsmarki2011,Narita2016,langenbach_2019,Sirenko2019,Sirenko2021}   but sometimes extending to the optical
region and beyond,\cite{Jung2004,Saito2008a,Saito2008b,Toyoda2015,Saito2008c,Saito2009,Yokosuk2020} and it arises from the fact that counter-propagating beams can have different absorption coefficients. 
\cite{Szaller2013,Kezsmarki2014,Cheong2018,Tokura2018,Cheong2019} 
In other words, a 
sample may be highly transmitting when measured with light
 in the +\textbf{k} direction but nearly opaque in the
-\textbf{k}   direction.
%
%
Nonreciprocity has a number of symmetry prerequisites. \cite{Cheong2018,Cheong2019} For a given light propagation direction \textbf{$\vec{\textrm k}$}, all
symmetries that reverse +\textbf{k} to -\textbf{k}
while leaving the sample unchanged 
must be broken including
inversion, mirrors or $C_2$ rotations about a plane or axis
perpendicular to \textbf{$\vec{\textrm k}$}, and time-reversal.\cite{Szaller2013} 
Several measurement configurations can be used to
take advantage of various types of symmetry breaking.\cite{Cheong2018,Cheong2019} Toroidal
dichroism occurs when light propagation
is along the toroidal moment  \textbf{T}  (\textbf{k} $\parallel$ \textbf{T} = \textbf{P}
$\times$ \textbf{{M}}, where \textbf{P} and \textbf{M} are the electric polarization and
magnetic moment).\cite{Kezsmarki2011,Bordacs2015,Rikken2002,Fishman2015,Narita2016,Ding2021} 
Magnetochiral dichroism, on the other hand, involves in both chirality and applied magnetic field.\cite{Bordacs2012,Saito2008a,Saito2008b,Rikken1997,Sessoli2015,Nakagawa2017,Train2011,Barron2008,Train2008,Sakano2020} 
%
Interestingly, nonreciprocity can be realized by reversing the external magnetic
field  as well as the light propagation direction. Linearly- and circularly-polarized light offer additional degrees of freedom,\cite{Narita2016} and with a vortex beam, one can even select specific projections of the orbital angular momentum.\cite{Sirenko2019,Sirenko2021} The spectroscopic response of these eigenstates is not always equal and opposite, leading to the phenomenon of nonreciprocal directional anisotropy.  As a result, unpolarized light can reveal these effects as well. \cite{Saito2008b,Yu2018,Yokosuk2020}

Because nonreciprocal directional dichroism requires cross-coupling between electric and magnetic dipole transitions, 
magnetoelectric
multiferroics - with their low crystallographic and magnetic symmetries - are promising platforms with which to search for these
effects.\cite{Kezsmarki2014,Yokosuk2020}  
%
Ni$_3$TeO$_6$ is a prominent example.  In addition to natural magnetoelectric character in the vicinity of the antiferromagnetic resonance,\cite{Sirenko2019,Sirenko2021} the crystal field (or color band) excitations are magnetoelectric and support broad band nonreciprocal directional dichroism.\cite{Yokosuk2020}  
As a polar + chiral magnet, Ni$_3$TeO$_6$ has the potential 
to host nonreciprocal effects not just in the chiral configurations - which were explored in our prior work\cite{Yokosuk2020} - but also in the toroidal geometry. 
%
Here, 
 we focus on magneto-optical spectroscopy and first-principles-based simulations of nonreciprocity 
 in the toroidal configuration 
 in order to complete this fascinating series and, at the same time,  unravel 
the structure-property relationships relevant to this unique light-matter interaction. For instance, we find that the 
largest contrast is supported by the $^3$A$_{\rm 2g}$ $\rightarrow$ $^3$T$_{\rm 2g}$ on-site excitations near 1.1 eV due to the creation of Ni$^{2+}$ toroidal moments in this relatively narrow energy range - quite different from the magnetochrial and transverse magnetochiral mechanisms that provide broader but more modest   effects over the color band range. 
The phonon side bands that ride on top of these $d$-to-$d$  excitations are quantitatively assigned to particular phonon modes and shown to display nonreciprocal effects as well.  In a significant conceptual advance, we demonstrate  dichroic contrast across the telecom range (i.e. optical fiber communication wavelengths). Remarkably, the telecom wavelength range dovetails perfectly with the strongest nonreciprocal response of Ni$_3$TeO$_6$ in the toroidal geometry. 
This establishes a 
potentially significant application for nonreciprocal materials - beyond ferrites and the microwave regime - in which optical circulators or directional amplifiers can operate in the telecom range with low loss and without complicated sample fabrication.\cite{Bi2011,Manipatruni2009,Peng2014,Scheucher2016}   
These findings also open the door to the use of toroidal magnetoelectrics for 
optical signal processing and communication.\cite{pozar2011}


\section*{Results and Discussion}


\subsection*{Crystal structure, properties, and symmetry analysis}

The crystal structure of Ni$_3$TeO$_6$ is both polar and chiral as expected for an $R3$ space group. \cite{Zivkovic2010} Chains of distorted NiO$_6$ and TeO$_6$ octahedra lie along the $c$ direction, and the three different Ni centers are distinguished by their local environments.\cite{Wang2015} 
In this work we denote the threefold axis of Ni$_3$TeO$_6$ as the `chiral axis' of the compound.\cite{Wang2015,Hlinka2014}
%
Although Ni$_3$TeO$_6$ hosts interlocking ferroelectric and chiral domain patterns,\cite{Wang2015,Ye2017}  crystals can be polished to reveal a single chiral domain.\cite{Yokosuk2020} Below $T_{\rm N}$ = 53 K, the system displays a collinear antiferromagnetic ground state. 
For \textbf{H}{$\parallel$}$c$, there is a spin-flop transition at 9 T leading to a conical spin and a metamagnetic transition at 52 T - both of which are accompanied by magnetoelastic effects.\cite{Oh2014,Kim2015,Yokosuk2015, Lass2020} There are no magnetically-driven phase transitions when field is applied perpendicular to the $c$-axis.\cite{Kim2015} 
The optical properties of Ni$_3$TeO$_6$ consist of several different color bands comprised of on-site Ni$^{2+}$ $d$-to-$d$ excitations along with a charge gap near 2.6 eV.\cite{Yokosuk2016} These features are sensitive to the spin-flop and metamagnetic transitions, and because spin-orbit coupling endows the excitations with magnetoelectric character,  these phases support nonreciprocal directional dichroism.\cite{Yokosuk2020} Here, the  excited Ni state (with a hole in the $t_{\rm 2g}$ orbital) provides spin-orbit coupling on the order of 40 meV and assures that the matrix elements
containing polarization and magnetization are non-zero. \cite{Yokosuk2020} Significantly, the broad band nonreciprocal directional dichroism in the magnetochiral configuration of Ni$_3$TeO$_6$ can be modeled using a quantitative first-principles-derived formalism.\cite{Yokosuk2020}

\begin{figure*}[tbh]
\begin{minipage}{5.5in}
\includegraphics[width = 5.5in]{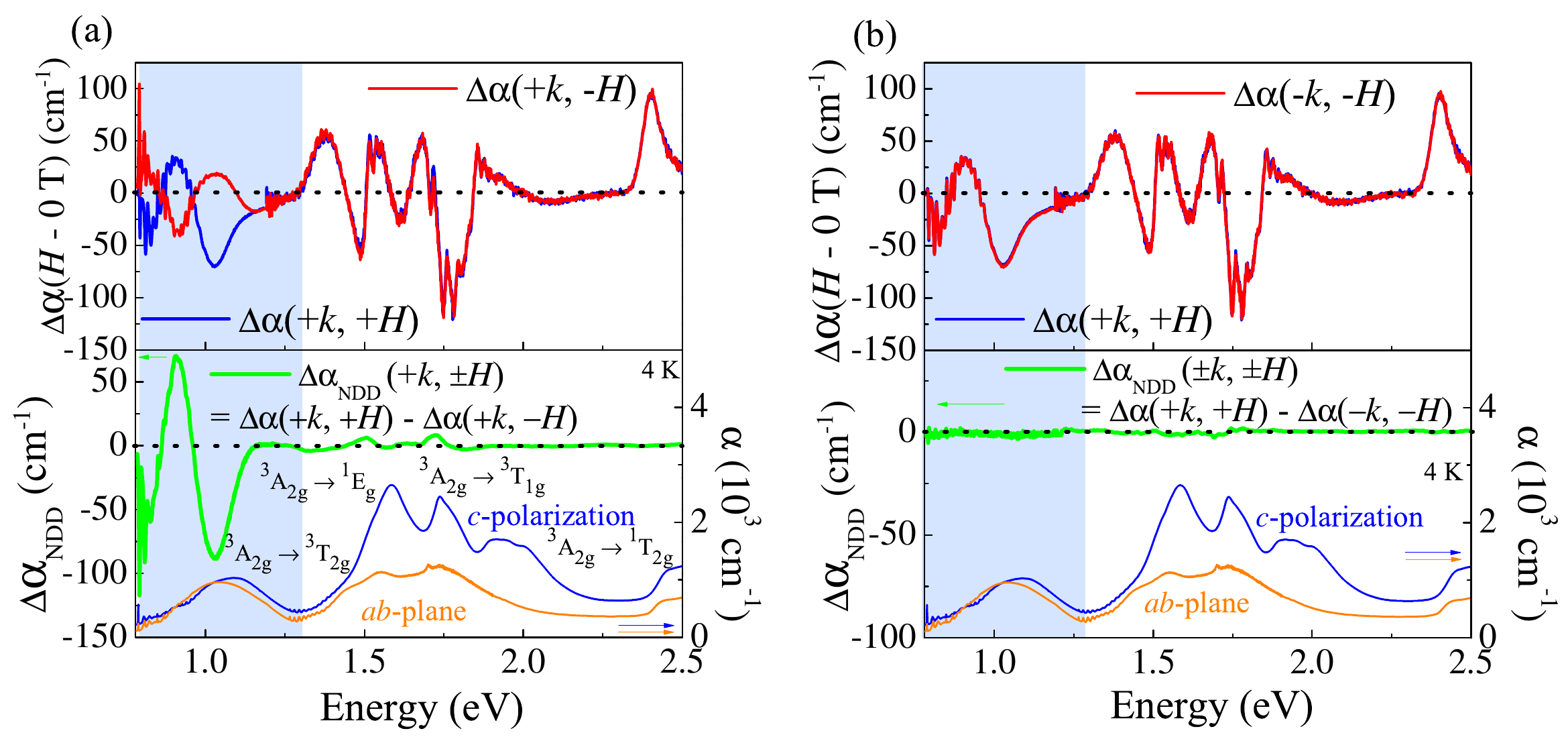}
\end{minipage}
\begin{minipage}{6.5in}
\caption{\label{Examples}
{\bf Illustrating symmetry considerations for nonreciprocity in Ni$_3$TeO$_6$.} (a) Full field absorption difference spectra of Ni$_3$TeO$_6$ as a function of energy for the (+\textbf{k}, +\textbf{H}) and (+\textbf{k}, -\textbf{H}) configurations in the toroidal measurement geometry. For illustration purposes, we employ \textbf{H} = 60 T.  Nonreciprocal directional dichroism  is shown in green.  Here,  $\Delta \alpha_{NDD}$(+\textbf{k}, $\pm$\textbf{H}) = $\Delta \alpha$(+\textbf{k}, +\textbf{H}) - $\Delta \alpha$(+\textbf{k}, -\textbf{H}). Experimentally, it is common to reverse the magnetic field direction as we have done here.  Nonreciprocity can also be defined in terms of counter-propagating beams, $\Delta \alpha_{NDD}$($\pm$\textbf{k}, +\textbf{H}), as  shown in Fig. 2.
(b)  Full field absorption difference spectra  in the (+\textbf{k}, +\textbf{H}) and (-\textbf{k}, -\textbf{H}) configurations.  Notice that they are the same.  ${\alpha}_{NDD}$($\pm$\textbf{k}, $\pm$\textbf{H}) = $\Delta \alpha$(+\textbf{k}, +\textbf{H}) - $\Delta \alpha$(-\textbf{k}, -\textbf{H})    (also shown in green)  vanishes when both light propagation and magnetic field direction are switched simultaneously. The absolute absorption spectrum is shown at the bottom of each panel to provide context. 
Note that we used cubic term symbols here ($^3$A$_{\rm 2g}$, $^3$T$_{\rm 2g}$, $^1$E$_{\rm g}$, $^3$T$_{\rm 1g}$,  $^1$T$_{\rm 2g}$) to denote Ni$^{2+}$ atomic multiplets because the deviation from cubic symmetry by trigonal distortion at the Ni sites is small ($\approx$0.1 eV in terms of crystal fields).
}
\end{minipage}
\end{figure*}

Figure \ref{Examples} summarizes our overall approach and the symmetry conditions that are important in this work. 
We begin by calculating absorption difference spectra  as   ${\Delta}{\alpha}$ = $\alpha$(\textbf{H}) - $\alpha$(0 T) for the different measurement configurations. 
Nonreciprocity is determined from various differences in the ${\Delta}{\alpha}$ spectra, for instance  $\Delta \alpha_{NDD}$(+\textbf{k}, $\pm$\textbf{H}) = $\Delta \alpha$(+\textbf{k}, +\textbf{H}) - $\Delta \alpha$(+\textbf{k}, -\textbf{H}) [Fig. \ref{Examples}(a)]. Notice that when both \textbf{k} and \textbf{H} are reversed,  ${\alpha}_{NDD}$($\pm$\textbf{k}, $\pm$\textbf{H}) = $\Delta \alpha$(+\textbf{k}, +\textbf{H}) - $\Delta \alpha$(-\textbf{k}, -\textbf{H}) vanishes [Fig. \ref{Examples}(b)]. Nonreciprocity can also be defined in terms of counter-propagating beams, $\Delta \alpha_{NDD}$($\pm$\textbf{k}, +\textbf{H}), as  shown in Fig. 2.

The top panels in Fig. \ref{NDD} summarize the three different measurement configurations predicted within the framework  of symmetry operational similarity. 
\cite{Cheong2018,Cheong2019}  In the toroidal configuration [Fig. \ref{NDD}(a)],  electric polarization along $c$, combined with \textbf{H} perpendicular to $c$, breaks two-fold rotation, inversion, mirror, and time-reversal symmetries, so it has symmetry operational similarity with the  wave vector of light \textbf{k} along the third direction. Thus, light propagation parallel to the toroidal moment can exhibit nonreciprocal directional dichroism. The two other configurations of interest involve chirality (rather than electric polarization) which breaks inversion and mirror symmetries. The magnetochiral geometry is obtained when the light propagation direction and applied field are parallel to the chiral axis [Fig. \ref{NDD}(b)]. The transverse magnetochiral configuration, on the other hand, requires both the light propagation direction and magnetic field to be perpendicular to the chiral axis [Fig. \ref{NDD}(c)]. 
Note that we denote the $C_3$ axis in Ni$_3$TeO$_6$ as the chiral axis; previous studies have argued that nonreciprocity in chiral systems can be easily understood considering chirality as a vector-like quantity.\cite{Hlinka2014,Cheong2018}

%
%

\subsection*{Nonreciprocal effect in different measurement configurations}

\begin{figure*}[tbh]
\begin{minipage}{6.5in}
\includegraphics[width = 6.5in]{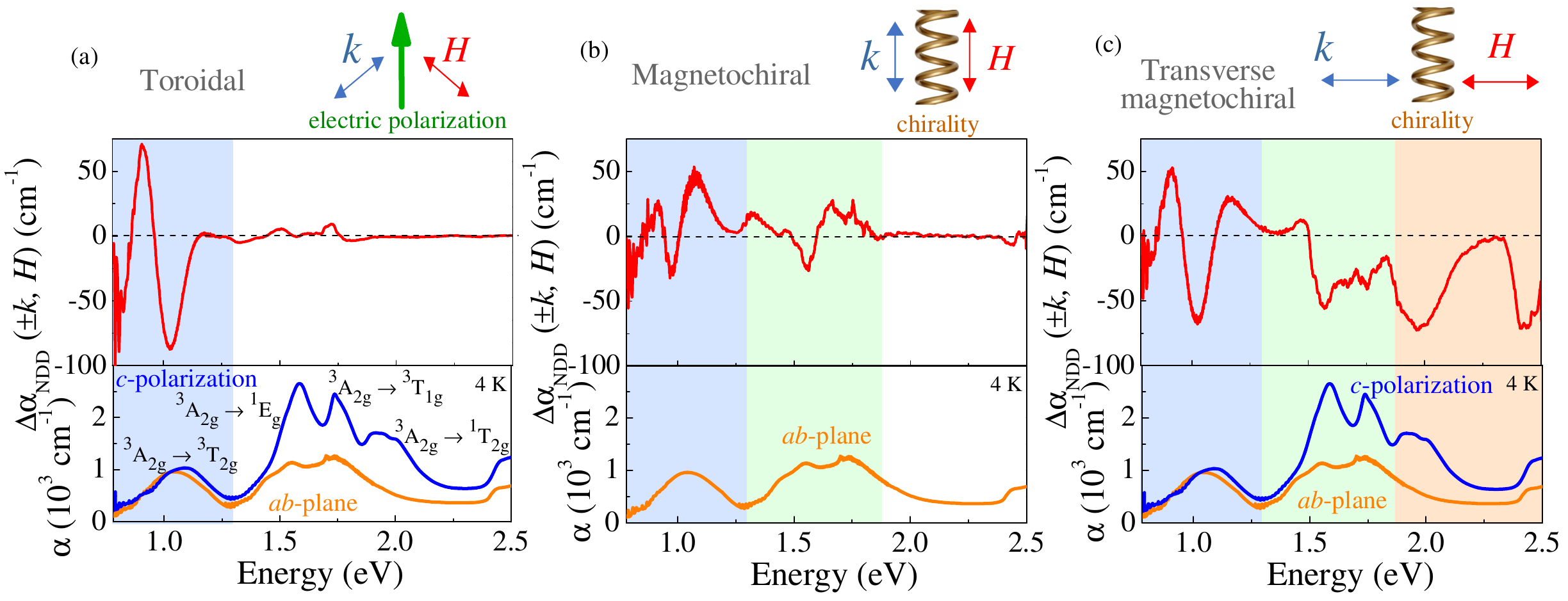}
\end{minipage}
\begin{minipage}{6.5in}
\caption{\label{NDD} {\bf Nonreciprocity of Ni$_3$TeO$_6$ in the three different measurement configurations.} Summary of the three different measurement configurations of interest in this work: (a) toroidal (\textbf{k} $\perp$ \textbf{H} $\perp$ polarization, (b) magnetochiral (\textbf{k} $\parallel$ \textbf{H} $\parallel$ chirality), and (c) transverse magnetochiral (\textbf{k} $\parallel$ \textbf{H} $\perp$ chirality).
Nonreciprocal directional dichroism spectra of Ni$_3$TeO$_6$ at 60 T in the toroidal (d), magnetochiral (e), and transverse magnetochiral (f) configurations, respectively. These spectra were measured in the Voigt (d) and Faraday (e, f) geometries. The nonreciprocal responses in panels (b, c) are reproduced from Ref. \citenum{Yokosuk2020}. 
The linear absorption spectrum of Ni$_3$TeO$_6$ at 4.2 K is shown below each nonreciprocal panel for comparison. The polarization directions and on-site Ni $d$-to-$d$ excitations are labeled. 
}
\end{minipage}
\end{figure*}

Figure \ref{NDD} summarizes nonreciprocal directional dichroism of Ni$_3$TeO$_6$ at full field in the toroidal, magnetochiral, and transverse magnetochiral configurations. 
Overall, this behavior is a consequence of low  symmetry, structural and magnetic chirality, and the presence of 
spin-orbit coupling, although the specific appearance of ${\Delta}{\alpha}_{\rm NDD}$ depends upon the measurement configuration as well. 
Focusing first on Ni$_3$TeO$_6$ in the toroidal geometry [Fig. \ref{NDD}(a)], we find a strong dichroic response in the vicinity of the 1.1 eV color band. This feature is assigned as a 
superposition of Ni$^{2+}$ on-site $d$-to-$d$  excitations ($^{3}$A$_{\rm 2g}$ $\rightarrow$ $^{3}$T$_{\rm 2g}$)  emanating from the three  different local environments of the Ni centers. \cite{Yokosuk2016}  At 60 T, the largest field-induced changes are centered on the 1.1 eV color band  and are on the order of 160 cm$^{-1}$. This corresponds to between zero and approximately 45\% contrast, depending on the value of the absolute absorption. 
The dichroic contrast in the vicinity of the $^{3}$A$_{\rm 2g}$ $\rightarrow$ $^{1}$E$_{\rm g}$ and $^{3}$A$_{\rm 2g}$ $\rightarrow$ $^{3}$T$_{\rm 1g}$ color bands is significantly smaller, and that at higher energy near the $^{3}$A$_{\rm g}$ $\rightarrow$ $^{1}$T$_{\rm 2g}$ excitation is zero within our sensitivity. In the toroidal configuration, the overall size of ${\Delta}{\alpha}_{\rm NDD}$ and its energy distribution  is quite different than what is observed in the magnetochiral and transverse magnetochiral geometries. \cite{Yokosuk2020}   For instance, while all configurations  exhibit dichroic contrast near the 1.1 eV color band, that in the toroidal configuration is by far the largest.

Turning to nonreciprocal directional dichroism of Ni$_3$TeO$_6$ in the magnetochiral configuration [Fig. \ref{NDD}(b)], ${\Delta}{\alpha}_{\rm NDD}$ associated with the 1.1 and 1.65 eV color bands is on the order of $\pm$45 cm$^{-1}$ at full field. \cite{Yokosuk2020}  This is different than what we find in the toroidal configuration both in terms of size and shape. There is a lot more fine structure to ${\Delta}{\alpha}_{\rm NDD}$ in the magnetochiral configuration as well. 
Unlike the other two measurement configurations shown in Fig. \ref{NDD}, ${\Delta}{\alpha}_{\rm NDD}$ in the magnetochiral geometry 
appears only above the 9 T spin flop transition where spin canting 
%
%
introduces net magnetization.\cite{Oh2014,Yokosuk2016}
The nonreciprocal response of Ni$_3$TeO$_6$ is different yet again in the transverse magnetochiral configuration with dichroic contrast associated with all four types of $d$-to-$d$ excitations in the absolute absorption spectrum [Fig. \ref{NDD}(c)]. This suggests that the color bands have magnetoelectric character in the chiral geometries as well. 
Differences in magnetoelectric coupling as it pertains to the various contrasts in toroidal, magnetochiral, and transverse magnetochiral geometries are discussed below.

\subsection*{Field dependence of the dichroic contrast in the toroidal geometry}

\begin{figure*}[tbh]
\begin{minipage}{5.2in}
\includegraphics[width = 5.2in]{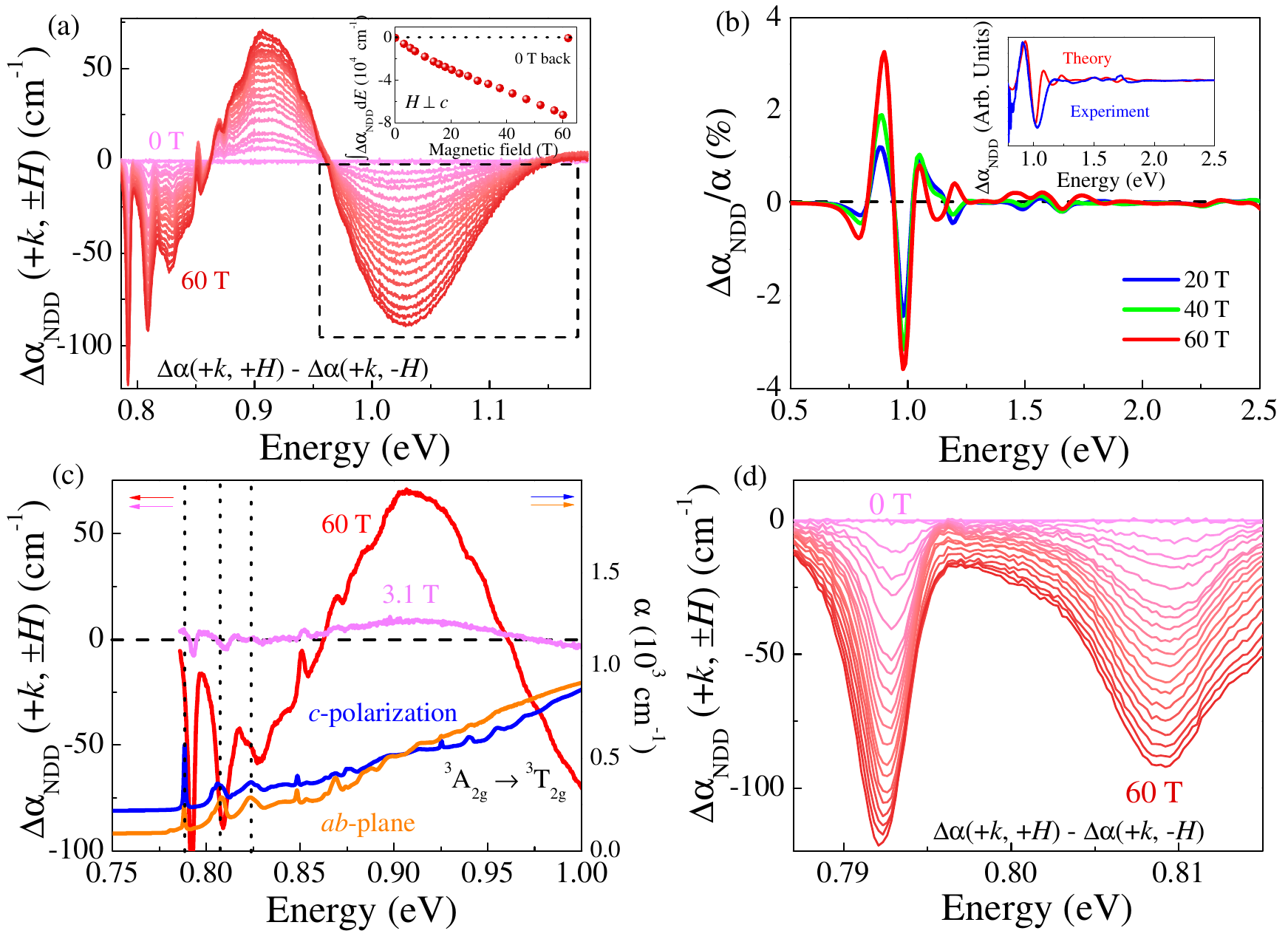}
\end{minipage}
\begin{minipage}{6.5in}
\caption{\label{NDD2} {\bf Nonreciprocity in the toroidal geometry.} (a) Nonreciprocal directional dichroism of Ni$_3$TeO$_6$ in the toroidal configuration measured between 0 and 60 T at 4.0 K. The inset is the integration of the indicated feature in the dashed box over an appropriate energy window. No hysteresis is observed. (b) Simulated nonreciprocal directional dichroism of Ni$_3$TeO$_6$ in the toroidal configuration.  The inset shows a comparison 
of the calculated and measured results at 60 T. The intensity is normalized to the maximum of ${\Delta}{\alpha}_{NDD}$. 
(c) Close-up view of  ${\Delta}{\alpha}_{\rm NDD}$ at 3.1 and 60 T compared with the linear absorption spectrum. (d) Close-up view of ${\Delta}{\alpha}_{\rm NDD}$ as a function of field between 0 and 60 T.  The complete list of magnetic fields in panels (a) and (d) is: 0, 3.1, 5.5, 7.2, 10.7, 13.8, 15.8, 17.9, 20.3, 23.1, 26.1, 29.6, 33.3, 37.5, 42.1, 47.0, 52.1, 57.1, 60.2 T.
}
\end{minipage}
\end{figure*}

Figure \ref{NDD2}(a) displays nonreciprocal directional dichroism of Ni$_3$TeO$_6$  as a function of magnetic field in the toroidal congifuration. As discussed previously, 
the toroidal geometry involves placing the electric polarization direction mutually orthogonal to both the light propagation and magnetic field directions. 
In addition to large positive and negative lobes in the dichroic contrast, there is a great deal of fine structure below 0.9 eV. These features
can be assigned to phonon sidebands as discussed below. 
Examination reveals that  $\alpha_{\rm NDD}$  appears at the lowest fields and grows systematically. 
This is because field-induced canting of Ni moments, which gives rise to net ferromagnetic moments and enables nonreciprocal behavior, can occur even at the smallest fields.
That nonreciprocal directional dichroism can be seen at low fields is useful for a number of applications including optical isolators and rectifiers, high-fidelity holograms, and potentially in the telecom sector. 
 In any case, we can quantify this trend  
by integrating $\alpha_{\rm NDD}$ over an appropriate energy window and plotting the result as a function of applied field [inset, Fig. \ref{NDD2}(a)]. The shape reveals how the spins align in the direction of the applied field. The lack of sharp jumps 
or cusps is consistent with the absence of field-induced magnetic transitions 
for \textbf{H} $\perp$ $c$. \cite{Kim2015}  No hysteresis is observed. In addition to switching the applied field direction, we  investigated symmetry effects \cite{Hlinka2014} in Ni$_3$TeO$_6$ by switching the light propagation direction [Supplementary Fig. 1]. The results are identical.  If both light propagation  and the magnetic field directions are switched at the same time, nonreciprocal directional dichroism vanishes due to time-reversal symmetry [Fig. \ref{Examples}(b) and Supplementary Fig. 2]. \cite{Hlinka2014}

%
%

Comparison between the experimental nonreciprocal directional dichroism in different geometries with the simulated versions provides insight into the nature of the  contrast in different energy ranges. 
Figure \ref{NDD2}(b) displays ${\Delta}\alpha_{\rm NDD}$ of Ni$_3$TeO$_6$ calculated by first-principles-based methods in the toroidal configuration.\cite{Yokosuk2020} The simulated spectrum compares reasonably well with the experimental result, except for the phonon sideband effects around 0.8 eV which are not included in the model, capturing the maximum near 0.9 eV, the minimum near 1.03 eV as well as the lack of contrast at higher energies.
Note that our simulation is based upon a local ionic picture and includes only intra-Ni-ionic excitations,\cite{Yokosuk2020} 
 not the entire part of the structural (and magnetic) chirality of Ni$_3$TeO$_6$. 
The structural chirality of Ni$_3$TeO$_6$ manifests (i) at the local level with a chiral crystal-field environment at each Ni site as well as (ii) in the global arrangement of Ni sites.  Our theory includes the former but not the latter. We refer to these as local and global chirality, respectively.\cite{Zhou2021}
Based on the results from our ionic picture-based simulation that nicely capture the dichroic response near 1 eV, 
we speculate that excitations near 1 eV are relevant to the local chiral component, namely 
the formation of the Ni ionic toroidal moments perpendicular to the magnetic field and bulk electric polarization 
($\textbf{T} \equiv \textbf{P} \times \textbf{M}$),
which are coupled with the propagation of light and induce nonreciprocal directional dichroism.\cite{Hlinka2014} 
On the other hand, nonreciprocity at higher energies, which is sizable only in the magnetochiral and transverse magnetochiral cases [Fig.~\ref{NDD}(b,c)], originates primarily from global chirality
and is difficult to capture within our atomistic simulation framework. Since our simulated nonreciprocal directional dichroism is given as a sum of individual Ni atomic contributions, some aspects of global chirality originating from arrangements of Ni sites at different Wyckoff positions may partially cancel out to yield smaller responses compared to experimental results.
Details are available in the Methods section and in Supplementary information [Supplementary Notes 1.3].
%

Antiferromagnets traditionally offer foundational opportunities to investigate collective excitations that arise from charge-spin and charge-lattice coupling. \cite{Sell1967} These features include excitons, zero phonon lines, phonon side bands, and magnon sidebands. They are commonly observed   on
the leading edge of the lowest energy $d$-to-$d$ band in antiferromagnets like MnF$_2$ and $\alpha$-Fe$_2$O$_3$. \cite{Greene1965,Chen2011} 
The absorption spectrum of Ni$_3$TeO$_6$ does not have a zero phonon line because it is in the strong coupling limit, \cite{Henderson1989} but it does have a rich set of phonon sidebands on the leading edge of the $^{3}$A$_{\rm 2g}$ $\rightarrow$ $^{3}$T$_{\rm 2g}$ excitation [Fig. \ref{NDD2}(c), Supplementary Fig. 6]. \cite{Yokosuk2020} These structures, often called `phonon replicas',  arise from  vibronic coupling, the details of which are characterized by the Huang-Rys factor. \cite{Henderson1989,Huang1950}
That the phonon progression on the leading edge of the Ni$^{2+}$ $d$-to-$d$ excitations matches features in ${\Delta}{\alpha}_{\rm NDD}$ illustrates the physical connection and demonstrates that the phonon sidebands have magnetoelectric character.
Nonreciprocal directional dichroism thus provides an opportunity to investigate how phonons activated in this manner are impacted by reversing 
the field and/or light propagation direction. At this time, there are only a handful of cases (BiFeO$_3$, few-layer CrI$_3$ and Cu$_2$OSeO$_3$) where phonons or phonon-derived properties such as thermal conductivity have been shown to host nonreciprocal effects.\cite{Liu2020,Nomura2019,Lee2016,Fishman2015} Nonreciprocal phonons obviously enable rectification of heat and sound\cite{Nomura2019} - a topic of sustained interest.
A giant nonreciprocal response in the THz range has recently been reported in Ni$_3$TeO$_6$,\cite{langenbach_2019} possibly originating from the effects mentioned above.  Below, we demonstrate that the phonon sidebands in Ni$_3$TeO$_6$ not only display nonreciprocal directional dichroism [Fig. \ref{NDD2}(d)] but that this  contrast occurs along with that related to the low-energy Ni$^{2+}$ crystal field excitations across the telecom range.

\subsection*{Nonreciprocal directional dichroism at telecom wavelengths}

We therefore return to the  spectra in Fig. \ref{NDD2}(c) which are even more intriguing when 
overlaid  with the useful telecom bands. The latter include the O-, E-, S-, C-, L- and U-bands.  Optical fiber communications typically operate in one of these windows. The C-band (1530 - 1565 nm) is perhaps most familiar and employs drawn Er-glass fibers. While the optical properties of Er-glass fibers have been studied separately under electric and magnetic fields, \cite{Leung1991,Guan2010,Wen2012,Nascimento2015,Zhang2017} we are unaware of any attempt to explore rectification effects.  Moreover, nonreciprocal directional dichroism has been revealed in a number of different energy regimes, it 
 is rarely encountered in 
the near infrared. \cite{Nakagawa2017,Sheu2019}

\begin{figure*}[tbh]
\begin{minipage}{4.0in}
\includegraphics[width = 4.0in]{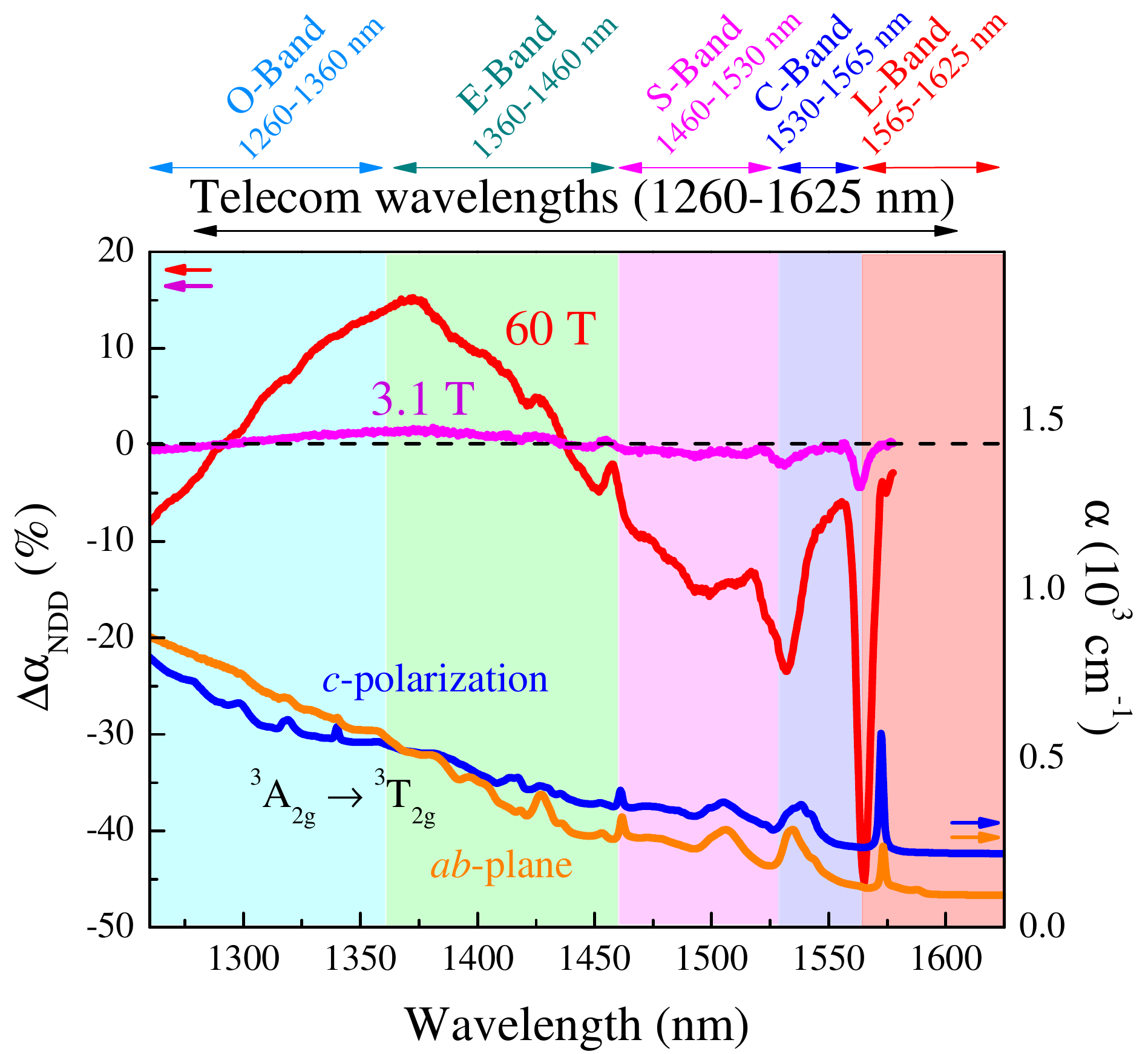}
\end{minipage}
\begin{minipage}{6.5in}
\caption{\label{Telecom} {\bf Nonreciprocity at telecom wavelengths.} Close-up view of the nonreciprocal directional dichroism in Ni$_3$TeO$_6$ on a percentage basis as a function of wavelength in the toroidal configuration at 4.0 K. ${\Delta}{\alpha}_{\rm NDD}$ is shown at 3.1 and 60 T.    The absolute absorption spectrum is included for comparison. Various  commercial telecom windows are summarized on the upper axes. The U-band (1625-1675 nm) is not shown.}
\end{minipage}
\end{figure*}

Figure  \ref{Telecom} displays a close-up view of the nonreciprocal directional dichroism in Ni$_3$TeO$_6$ in the toroidal configuration plotted on a percentage basis and a wavelength scale. Examination reveals that ${\Delta}{\alpha}_{\rm NDD}$ has a near perfect match to the full range of telecom wavelengths. The effects are strong, broad, and clearly evident at full field (60 T) with a maximum contrast in the first phonon side band of approximately 45\%.  These features also appear even at the smallest fields (as shown by the spectrum at 3.1 T) because time-reversal is already broken. This, along with the fact that we are using unpolarized light, is an advantage for applications. Difference signals are already in regular use, and nonreciprocal effects can add an important layer of security. Ni$_3$TeO$_6$ hosts similar contrast in the magnetochiral and transverse magnetochiral configurations as well [Supplementary Fig. 7].  In addition to demonstrating that nonreciprocal directional dichroism can be positioned within useful telecom windows, this work opens the door to the development of  high efficiency / low dissipation optical diodes and rectifiers from crystalline materials. 
 We anticipate that linear or circular polarizers would amplify the size of this effect in Ni$_3$TeO$_6$, but the ability to achieve polarization-independent signal is one of the beauties of magnetochiral materials.

\subsection*{Structure-property relations for photonics applications}

Ni$_3$TeO$_6$ is a superb platform for fundamental studies of nonreciprocity because it hosts this peculiar property across a wide range of excitations and in three different measurement configurations. To our knowledge, there are no other nonreciprocal materials that have been studied in so many different geometries. 
In this work, we focus on the Ni$^{2+}$ $d$-to-$d$ excitations and associated phonon sidebands in the near infrared and optical range in the toroidal configuration, unraveling structure-property relationships via comparison with contrast in the magnetochiral and transverse magnetochiral geometries.
We find that polarity allows the creation of Ni toroidal moments, which results in large contrast near the color band at 1.1 eV.  The chiral geometries, on the other hand, yield smaller contrast over a much broader energy range. 
Our modeling of the size and shape of the spectral response supports this picture of the importance of polarity vs. chirality in the creation of large vs. broadband contrast. These findings enhance our ability to design and deliver complex materials properties on demand. At the same time, the discovery of nonreciprocal effects across the full telecom wavelength range 
presents a number of exciting opportunities. While tunability under magnetic field is established in this work, the degree to which nonreciprocity can be controlled by other external stimuli or chemical substitution is unexplored. That said, the discovery 
of nonreciprocity in the telecom region is a significant conceptual advance that has the potential to jump-start the use of dichroic contrast in photonics applications -  
particularly in the area of high-efficiency optical diodes and rectifiers. 


\section*{Methods}


\paragraph*{\bf Crystal growth and orientation:} High quality single crystals of Ni$_3$TeO$_6$ were grown by chemical vapour transport methods as described previously.\cite{Oh2014} The crystals were polished to expose either the $ab$-plane or the $c$-axis and to control optical density. After polishing, the  sample thicknesses were  on the order of 30 $\mu$m. Optical microscope images and optical rotation data confirm that the crystals have a single chiral domain.\cite{Yokosuk2020} 
The crystals were coated with transparent epoxy to stabilize the structure during the field pulses. 

\paragraph*{\bf Optical spectroscopy:} 
Polarized optical transmittance was measured as a function of energy and temperature using a series of spectrometers as described previously (0.78 - 2.5 eV; 4.2  -300 K). \cite{Yokosuk2015} Absorption was calculated as $\alpha (E)$ = $\frac{1}{d}$ln($T(E)$), where $d$ is the sample thickness and $T(E)$ is the measured transmittance.

\paragraph*{\bf Magneto-optical spectroscopy:} 
Magneto-optical work was performed in the toroidal geometry in a capacitor-driven 65 T pulsed magnet at the National High Magnetic Field Laboratory in Los Alamos, NM. We employed the standard transmission probe fitted with a specially-designed Voigt end-piece for this work.
These measurements covered the 0.75 - 2.6 eV range with 2.4 meV resolution and were carried out at 4 K. 
Broadband light from a tungsten lamp was coupled to optical fibers and focused onto the sample for transmittance experiments. 
A collection fiber brought the light from the top of the probe to the grating spectrometer, where both CCD and InGaAs detectors were employed as appropriate. 
The spectra were taken in four different measurement configurations: ($\pm$\textbf{k}, $\pm$\textbf{H}).
Each run was carried out sequentially and consistently, starting with one \textbf{k} direction (and pulsing to obtained both ${\pm}$\textbf{H}) and then switching to the other \textbf{k} direction  by swapping the fibers (again  pulsing both up and down). 
%
We calculated the absorption differences as: $\Delta\alpha$ = $\alpha$(\textbf{H} = $\pm$ 60 T) - $\alpha$(\textbf{H} = 0 T). As an example, nonreciprocal directional dichroism is calculated as:  $\Delta\alpha_{\rm NDD}$(+\textbf{k}, $\pm$ \textbf{H}) = $\Delta\alpha$(+\textbf{k}, +\textbf{H}) - $\Delta\alpha$(+\textbf{k}, -\textbf{H}). 
Nonreciprocity can also be defined in terms of counter-propagating beams as  $\Delta \alpha_{NDD}$($\pm$\textbf{k}, +\textbf{H}). 
Traditional smoothing techniques were 
employed in the CCD detector regime.

\paragraph*{\bf Modeling the optical diode effect:} 

Magnetic configurations under the external \textbf{H}-field were found by optimizing total energies of a magnetic exchange hamiltonian derived from previous first-principles electronic structure calculations.\cite{Kim2015} Distortions in the crystal structure induced by the external \textbf{H}-field were simulated by enforcing magnetic configurations obtained from the model optimization above during the first-principles density functional theory (DFT) calculations and structural relaxations, which were performed via the Vienna {\it ab-initio} Simulation Package ({\sc vasp}).\cite{VASP1,VASP2} For {\sc vasp} structural relaxations we employed a 500 eV of plane-wave energy cutoff, a $5 \times 5 \times 5$ $\Gamma$-centered \textbf{k}-point sampling, and a simplified rotationally-invariant DFT+$U_{\rm eff}$ formalism ($U_{\rm eff} = 4$ eV)\cite{Dudarev1998} on top of PBEsol exchange-correlation functional.\cite{PBEsol} After structural relaxations, {\sc wien2k} full-potential DFT code\cite{wien2k} was used to compute electric dipole matrix elements and crystal fields in terms of the non-interacting band basis, which were then projected onto Ni$^{2+}$  atomic multiplet states via exact diagonalization (ED) routine included in the Embedded DMFT Functional ({\sc edmftf}) code.\cite{eDMFT} Details on {\sc wien2k}, ED calculations, computations of electric and magnetic dipole matrix elements, and estimations of magneto-electric response tensors are presented in our previous work.\cite{Yokosuk2020} 


\section*{Data Availability}

Data are available from the corresponding author upon reasonable request.

\section*{Code Availability}

The codes implementing the calculations of this study are available from the corresponding author upon request.

\section*{Acknowledgements}

Research at the University of Tennessee and Rutgers University is supported by the NSF-DMREF 
 program (DMR-1629079 and DMR-1629059). 
 A portion of this work was performed at the National High Magnetic Field Laboratory which is supported by the National Science Foundation DMR-1644779, the State of Florida, and the U.S. Department of Energy.
 H.-S.K. acknowledges funding from the Basic Science Research Program through the National Research Foundation of Korea funded by the Ministry of Education (NRF-2020R1C1C1005900), and also the support of computational resources, including technical assistance from the National Supercomputing Center of Korea (Grant No. KSC-2020-CRE-0156).
 We thank B. Donahoe for useful conversations.

\section*{Author Contributions}

KP, MOY, and JLM designed the study. JY and SWC grew the crystals. KP, MOY, and JLM discussed the measurement configurations and run pattern in detail. KP, MOY, MG, and SAC performed the pulsed field optical measurements, and KP and JLM analyzed the spectral data. HSK, KH, and DV developed a microscopic model for nonreciprocal optical effects and applied it to Ni$_3$TeO$_6$. KP, HSK, and JLM wrote the manuscript. All authors commented on the text.

\section*{Competing Interests}

The authors declare no competing interests

\bibliographystyle{naturemag}
\bibliography{citation}

\begin{thebibliography}{10}
\expandafter\ifx\csname url\endcsname\relax
  \def\url#1{\texttt{#1}}\fi
\expandafter\ifx\csname urlprefix\endcsname\relax\def\urlprefix{URL }\fi
\providecommand{\bibinfo}[2]{#2}
\providecommand{\eprint}[2][]{\url{#2}}

\bibitem{Kezsmarki2014}
\bibinfo{author}{K{\'{e}}zsm{\'{a}}rki, I.} \emph{et~al.}
\newblock \bibinfo{title}{One-way transparency of four-coloured spin-wave
  excitations in multiferroic materials}.
\newblock \emph{\bibinfo{journal}{Nat. Commun.}} \textbf{\bibinfo{volume}{5}},
  \bibinfo{pages}{3203} (\bibinfo{year}{2014}).

\bibitem{Tokura2018}
\bibinfo{author}{Tokura, Y.} \& \bibinfo{author}{Nagaosa, N.}
\newblock \bibinfo{title}{Nonreciprocal responses from non-centrosymmetric
  quantum materials}.
\newblock \emph{\bibinfo{journal}{Nat. Commun.}} \textbf{\bibinfo{volume}{9}},
  \bibinfo{pages}{3740} (\bibinfo{year}{2018}).

\bibitem{Kezsmarki2015}
\bibinfo{author}{K\'ezsm\'arki, I.} \emph{et~al.}
\newblock \bibinfo{title}{{Optical diode effect at spin-wave excitations of the
  room-temperature multiferroic BiFeO$_3$}}.
\newblock \emph{\bibinfo{journal}{Phys. Rev. Lett.}}
  \textbf{\bibinfo{volume}{115}}, \bibinfo{pages}{127203}
  (\bibinfo{year}{2015}).

\bibitem{Lee2016}
\bibinfo{author}{Lee, J.~H.}, \bibinfo{author}{K{\'{e}}zsm{\'{a}}ki, I.} \&
  \bibinfo{author}{Fishman, R.~S.}
\newblock \bibinfo{title}{{First-principles approach to the dynamic
  magnetoelectric couplings for the non-reciprocal directional dichroism in
  BiFeO$_3$}}.
\newblock \emph{\bibinfo{journal}{New J. Phys.}} \textbf{\bibinfo{volume}{18}},
  \bibinfo{pages}{043025} (\bibinfo{year}{2016}).

\bibitem{Bordacs2015}
\bibinfo{author}{Bord\'acs, S.} \emph{et~al.}
\newblock \bibinfo{title}{{Unidirectional terahertz light absorption in the
  pyroelectric ferrimagnet ${\mathrm{CaBaCo}}_{4}{\mathrm{O}}_{7}$}}.
\newblock \emph{\bibinfo{journal}{Phys. Rev. B}} \textbf{\bibinfo{volume}{92}},
  \bibinfo{pages}{214441} (\bibinfo{year}{2015}).

\bibitem{Yu2018}
\bibinfo{author}{Yu, S.} \emph{et~al.}
\newblock \bibinfo{title}{{High-temperature terahertz optical diode effect
  without magnetic order in polar ${\mathrm{FeZnMo}}_{3}{\mathrm{O}}_{8}$}}.
\newblock \emph{\bibinfo{journal}{Phys. Rev. Lett.}}
  \textbf{\bibinfo{volume}{120}}, \bibinfo{pages}{037601}
  (\bibinfo{year}{2018}).

\bibitem{Sheu2019}
\bibinfo{author}{Sheu, Y.~M.} \emph{et~al.}
\newblock \bibinfo{title}{{Picosecond creation of switchable optomagnets from a
  polar antiferromagnet with giant photoinduced Kerr rotations}}.
\newblock \emph{\bibinfo{journal}{Phys. Rev. X}} \textbf{\bibinfo{volume}{9}},
  \bibinfo{pages}{031038} (\bibinfo{year}{2019}).

\bibitem{Bordacs2012}
\bibinfo{author}{Bord{\'{a}}cs, S.} \emph{et~al.}
\newblock \bibinfo{title}{Chirality of matter shows up via spin excitations}.
\newblock \emph{\bibinfo{journal}{Nat. Phys.}} \textbf{\bibinfo{volume}{8}},
  \bibinfo{pages}{734--738} (\bibinfo{year}{2012}).

\bibitem{Kezsmarki2011}
\bibinfo{author}{K\'ezsm\'arki, I.} \emph{et~al.}
\newblock \bibinfo{title}{{Enhanced directional dichroism of terahertz light in
  resonance with magnetic excitations of the multiferroic Ba$_2$CoGe$_2$O$_7$
  oxide compound}}.
\newblock \emph{\bibinfo{journal}{Phys. Rev. Lett.}}
  \textbf{\bibinfo{volume}{106}}, \bibinfo{pages}{057403}
  (\bibinfo{year}{2011}).

\bibitem{Narita2016}
\bibinfo{author}{Narita, H.} \emph{et~al.}
\newblock \bibinfo{title}{{Observation of nonreciprocal directional dichroism
  via electromagnon resonance in a chiral-lattice helimagnet
  $\mathrm{B}{\mathrm{a}}_{3}\mathrm{NbF}{\mathrm{e}}_{3}\mathrm{S}{\mathrm{i}}_{2}{\mathrm{O}}_{14}$}}.
\newblock \emph{\bibinfo{journal}{Phys. Rev. B}} \textbf{\bibinfo{volume}{94}},
  \bibinfo{pages}{094433} (\bibinfo{year}{2016}).

\bibitem{langenbach_2019}
\bibinfo{author}{Langenbach, M.}
\newblock \emph{\bibinfo{title}{Giant directional dichroism in chiral
  Ni$_3$TeO$_6$ in THz spectroscopy in high magnetic fields}}.
\newblock Ph.D. thesis (\bibinfo{year}{2019}).
\newblock
  \urlprefix\url{https://www.semanticscholar.org/paper/Giant-directional-dichroism-in-chiral-Ni_3TeO_6-in-Langenbach/8fda74dc714b6c626d6b9897a958e69aa0e30495}.

\bibitem{Sirenko2019}
\bibinfo{author}{Sirenko, A.~A.} \emph{et~al.}
\newblock \bibinfo{title}{Terahertz vortex beam as a spectroscopic probe of
  magnetic excitations}.
\newblock \emph{\bibinfo{journal}{Phys. Rev. Lett.}}
  \textbf{\bibinfo{volume}{122}}, \bibinfo{pages}{237401}
  (\bibinfo{year}{2019}).

\bibitem{Sirenko2021}
\bibinfo{author}{Sirenko, A.~A.} \emph{et~al.}
\newblock \bibinfo{title}{Total angular momentum dichroism of the terahertz
  vortex beams at the antiferromagnetic resonances}.
\newblock \emph{\bibinfo{journal}{Phys. Rev. Lett.}}
  \textbf{\bibinfo{volume}{126}}, \bibinfo{pages}{157401}
  (\bibinfo{year}{2021}).

\bibitem{Jung2004}
\bibinfo{author}{Jung, J.~H.} \emph{et~al.}
\newblock \bibinfo{title}{{Optical magnetoelectric effect in the polar
  ${\mathrm{G}\mathrm{a}\mathrm{F}\mathrm{e}\mathrm{O}}_{3}$ ferrimagnet}}.
\newblock \emph{\bibinfo{journal}{Phys. Rev. Lett.}}
  \textbf{\bibinfo{volume}{93}}, \bibinfo{pages}{037403}
  (\bibinfo{year}{2004}).

\bibitem{Saito2008a}
\bibinfo{author}{Saito, M.}, \bibinfo{author}{Taniguchi, K.} \&
  \bibinfo{author}{Arima, T.}
\newblock \bibinfo{title}{{Gigantic optical magnetoelectric effect in
  CuB$_2$O$_4$}}.
\newblock \emph{\bibinfo{journal}{J. Phys. Soc. Jpn}}
  \textbf{\bibinfo{volume}{77}}, \bibinfo{pages}{013705}
  (\bibinfo{year}{2008}).

\bibitem{Saito2008b}
\bibinfo{author}{Saito, M.}, \bibinfo{author}{Ishikawa, K.},
  \bibinfo{author}{Taniguchi, K.} \& \bibinfo{author}{Arima, T.}
\newblock \bibinfo{title}{{Magnetic control of crystal chirality and the
  existence of a large magneto-optical dichroism effect in
  ${\mathrm{CuB}}_{2}{\mathrm{O}}_{4}$}}.
\newblock \emph{\bibinfo{journal}{Phys. Rev. Lett.}}
  \textbf{\bibinfo{volume}{101}}, \bibinfo{pages}{117402}
  (\bibinfo{year}{2008}).

\bibitem{Toyoda2015}
\bibinfo{author}{Toyoda, S.} \emph{et~al.}
\newblock \bibinfo{title}{{One-way transparency of light in multiferroic
  ${\mathrm{CuB}}_{2}{\mathrm{O}}_{4}$}}.
\newblock \emph{\bibinfo{journal}{Phys. Rev. Lett.}}
  \textbf{\bibinfo{volume}{115}}, \bibinfo{pages}{267207}
  (\bibinfo{year}{2015}).

\bibitem{Saito2008c}
\bibinfo{author}{Saito, M.}, \bibinfo{author}{Ishikawa, K.},
  \bibinfo{author}{Taniguchi, K.} \& \bibinfo{author}{Arima, T.}
\newblock \bibinfo{title}{{Magnetically controllable CuB$_2$O$_4$ phase
  retarder}}.
\newblock \emph{\bibinfo{journal}{Appl. Phys. Express}}
  \textbf{\bibinfo{volume}{1}}, \bibinfo{pages}{121302} (\bibinfo{year}{2008}).

\bibitem{Saito2009}
\bibinfo{author}{Saito, M.}, \bibinfo{author}{Ishikawa, K.},
  \bibinfo{author}{Konno, S.}, \bibinfo{author}{Taniguchi, K.} \&
  \bibinfo{author}{Arima, T.}
\newblock \bibinfo{title}{{Periodic rotation of magnetization in a
  non-centrosymmetric soft magnet induced by an electric field}}.
\newblock \emph{\bibinfo{journal}{Nat. Mater.}} \textbf{\bibinfo{volume}{8}},
  \bibinfo{pages}{634--638} (\bibinfo{year}{2009}).

\bibitem{Yokosuk2020}
\bibinfo{author}{Yokosuk, M.~O.} \emph{et~al.}
\newblock \bibinfo{title}{{Nonreciprocal directional dichroism of a chiral
  magnet in the visible range}}.
\newblock \emph{\bibinfo{journal}{npj Quant. Mater.}}
  \textbf{\bibinfo{volume}{5}}, \bibinfo{pages}{20} (\bibinfo{year}{2020}).

\bibitem{Szaller2013}
\bibinfo{author}{Szaller, D.}, \bibinfo{author}{Bord\'acs, S.} \&
  \bibinfo{author}{K\'ezsm\'arki, I.}
\newblock \bibinfo{title}{Symmetry conditions for nonreciprocal light
  propagation in magnetic crystals}.
\newblock \emph{\bibinfo{journal}{Phys. Rev. B}} \textbf{\bibinfo{volume}{87}},
  \bibinfo{pages}{014421} (\bibinfo{year}{2013}).

\bibitem{Cheong2018}
\bibinfo{author}{Cheong, S.-W.}, \bibinfo{author}{Talbayev, D.},
  \bibinfo{author}{Kiryukhin, V.} \& \bibinfo{author}{Saxena, A.}
\newblock \bibinfo{title}{Broken symmetries, non-reciprocity, and
  multiferroicity}.
\newblock \emph{\bibinfo{journal}{npj Quant. Mater.}}
  \textbf{\bibinfo{volume}{3}}, \bibinfo{pages}{19} (\bibinfo{year}{2018}).

\bibitem{Cheong2019}
\bibinfo{author}{Cheong, S.-W.}
\newblock \bibinfo{title}{{SOS}: symmetry-operational similarity}.
\newblock \emph{\bibinfo{journal}{npj Quant. Mater.}}
  \textbf{\bibinfo{volume}{4}}, \bibinfo{pages}{53} (\bibinfo{year}{2019}).

\bibitem{Rikken2002}
\bibinfo{author}{Rikken, G. L. J.~A.}, \bibinfo{author}{Strohm, C.} \&
  \bibinfo{author}{Wyder, P.}
\newblock \bibinfo{title}{Observation of magnetoelectric directional
  anisotropy}.
\newblock \emph{\bibinfo{journal}{Phys. Rev. Lett.}}
  \textbf{\bibinfo{volume}{89}}, \bibinfo{pages}{133005}
  (\bibinfo{year}{2002}).

\bibitem{Fishman2015}
\bibinfo{author}{Fishman, R.~S.} \emph{et~al.}
\newblock \bibinfo{title}{{Spin-induced polarizations and nonreciprocal
  directional dichroism of the room-temperature multiferroic
  ${\mathrm{BiFeO}}_{3}$}}.
\newblock \emph{\bibinfo{journal}{Phys. Rev. B}} \textbf{\bibinfo{volume}{92}},
  \bibinfo{pages}{094422} (\bibinfo{year}{2015}).

\bibitem{Ding2021}
\bibinfo{author}{Ding, L.} \emph{et~al.}
\newblock \bibinfo{title}{{Field-tunable toroidal moment in a chiral-lattice
  magnet}}.
\newblock \emph{\bibinfo{journal}{Nat. Commun.}} \textbf{\bibinfo{volume}{12}},
  \bibinfo{pages}{5339} (\bibinfo{year}{2021}).

\bibitem{Rikken1997}
\bibinfo{author}{Rikken, G. L. J.~A.} \& \bibinfo{author}{Raupach, E.}
\newblock \bibinfo{title}{Observation of magneto-chiral dichroism}.
\newblock \emph{\bibinfo{journal}{Nature}} \textbf{\bibinfo{volume}{390}},
  \bibinfo{pages}{493--494} (\bibinfo{year}{1997}).

\bibitem{Sessoli2015}
\bibinfo{author}{Sessoli, R.} \emph{et~al.}
\newblock \bibinfo{title}{{Strong magneto-chiral dichroism in a paramagnetic
  molecular helix observed by hard X-rays}}.
\newblock \emph{\bibinfo{journal}{Nat. Phys.}} \textbf{\bibinfo{volume}{11}},
  \bibinfo{pages}{69--74} (\bibinfo{year}{2015}).

\bibitem{Nakagawa2017}
\bibinfo{author}{Nakagawa, N.} \emph{et~al.}
\newblock \bibinfo{title}{{Magneto-chiral dichroism of
  ${\mathrm{CsCuCl}}_{3}$}}.
\newblock \emph{\bibinfo{journal}{Phys. Rev. B}} \textbf{\bibinfo{volume}{96}},
  \bibinfo{pages}{121102} (\bibinfo{year}{2017}).

\bibitem{Train2011}
\bibinfo{author}{Train, C.}, \bibinfo{author}{Gruselle, M.} \&
  \bibinfo{author}{Verdaguer, M.}
\newblock \bibinfo{title}{The fruitful introduction of chirality and control of
  absolute configurations in molecular magnets}.
\newblock \emph{\bibinfo{journal}{Chem. Soc. Rev.}}
  \textbf{\bibinfo{volume}{40}}, \bibinfo{pages}{3297} (\bibinfo{year}{2011}).

\bibitem{Barron2008}
\bibinfo{author}{Barron, L.~D.}
\newblock \bibinfo{title}{Chirality and magnetism shake hands}.
\newblock \emph{\bibinfo{journal}{Nat. Mater.}} \textbf{\bibinfo{volume}{7}},
  \bibinfo{pages}{691--692} (\bibinfo{year}{2008}).

\bibitem{Train2008}
\bibinfo{author}{Train, C.} \emph{et~al.}
\newblock \bibinfo{title}{Strong magneto-chiral dichroism in enantiopure chiral
  ferromagnets}.
\newblock \emph{\bibinfo{journal}{Nat. Mater.}} \textbf{\bibinfo{volume}{7}},
  \bibinfo{pages}{729--734} (\bibinfo{year}{2008}).

\bibitem{Sakano2020}
\bibinfo{author}{Sakano, M.} \emph{et~al.}
\newblock \bibinfo{title}{Radial spin texture in elemental tellurium with
  chiral crystal structure}.
\newblock \emph{\bibinfo{journal}{Phys. Rev. Lett.}}
  \textbf{\bibinfo{volume}{124}}, \bibinfo{pages}{136404}
  (\bibinfo{year}{2020}).

\bibitem{Bi2011}
\bibinfo{author}{Bi, L.} \emph{et~al.}
\newblock \bibinfo{title}{On-chip optical isolation in monolithically
  integrated non-reciprocal optical resonators}.
\newblock \emph{\bibinfo{journal}{Nat. Photon.}} \textbf{\bibinfo{volume}{5}},
  \bibinfo{pages}{758--762} (\bibinfo{year}{2011}).

\bibitem{Manipatruni2009}
\bibinfo{author}{Manipatruni, S.}, \bibinfo{author}{Robinson, J.~T.} \&
  \bibinfo{author}{Lipson, M.}
\newblock \bibinfo{title}{Optical nonreciprocity in optomechanical structures}.
\newblock \emph{\bibinfo{journal}{Phys. Rev. Lett.}}
  \textbf{\bibinfo{volume}{102}}, \bibinfo{pages}{213903}
  (\bibinfo{year}{2009}).

\bibitem{Peng2014}
\bibinfo{author}{Peng, B.} \emph{et~al.}
\newblock \bibinfo{title}{Parity--time-symmetric whispering-gallery
  microcavities}.
\newblock \emph{\bibinfo{journal}{Nat. Phys.}} \textbf{\bibinfo{volume}{10}},
  \bibinfo{pages}{394--398} (\bibinfo{year}{2014}).

\bibitem{Scheucher2016}
\bibinfo{author}{Scheucher, M.}, \bibinfo{author}{Hilico, A.},
  \bibinfo{author}{Will, E.}, \bibinfo{author}{Volz, J.} \&
  \bibinfo{author}{Rauschenbeutel, A.}
\newblock \bibinfo{title}{Quantum optical circulator controlled by a single
  chirally coupled atom}.
\newblock \emph{\bibinfo{journal}{Science}} \textbf{\bibinfo{volume}{354}},
  \bibinfo{pages}{1577--1580} (\bibinfo{year}{2016}).

\bibitem{pozar2011}
\bibinfo{author}{Pozar, D.~M.}
\newblock \emph{\bibinfo{title}{Microwave engineering}}
  (\bibinfo{publisher}{Wiley}, \bibinfo{year}{2011}).

\bibitem{Zivkovic2010}
\bibinfo{author}{{\v{Z}}ivkovi{\'{c}}, I.}, \bibinfo{author}{Pr{\v{s}}a, K.},
  \bibinfo{author}{Zaharko, O.} \& \bibinfo{author}{Berger, H.}
\newblock \bibinfo{title}{{Ni$_3$TeO$_6$ — a collinear antiferromagnet with
  ferromagnetic honeycomb planes}}.
\newblock \emph{\bibinfo{journal}{J. Phys.: Condens. Matter}}
  \textbf{\bibinfo{volume}{22}}, \bibinfo{pages}{056002}
  (\bibinfo{year}{2010}).

\bibitem{Wang2015}
\bibinfo{author}{Wang, X.}, \bibinfo{author}{Huang, F.-T.},
  \bibinfo{author}{Yang, J.}, \bibinfo{author}{Oh, Y.~S.} \&
  \bibinfo{author}{Cheong, S.-W.}
\newblock \bibinfo{title}{{Interlocked chiral/polar domain walls and large
  optical rotation in Ni$_3$TeO$_6$}}.
\newblock \emph{\bibinfo{journal}{APL Mater.}} \textbf{\bibinfo{volume}{3}},
  \bibinfo{pages}{076105} (\bibinfo{year}{2015}).

\bibitem{Hlinka2014}
\bibinfo{author}{Hlinka, J.}
\newblock \bibinfo{title}{Eight types of symmetrically distinct vectorlike
  physical quantities}.
\newblock \emph{\bibinfo{journal}{Phys. Rev. Lett.}}
  \textbf{\bibinfo{volume}{113}}, \bibinfo{pages}{165502}
  (\bibinfo{year}{2014}).

\bibitem{Ye2017}
\bibinfo{author}{Ye, M.} \& \bibinfo{author}{Vanderbilt, D.}
\newblock \bibinfo{title}{Domain walls and ferroelectric reversal in corundum
  derivatives}.
\newblock \emph{\bibinfo{journal}{Phys. Rev. B}} \textbf{\bibinfo{volume}{95}},
  \bibinfo{pages}{014105} (\bibinfo{year}{2017}).

\bibitem{Oh2014}
\bibinfo{author}{Oh, Y.~S.} \emph{et~al.}
\newblock \bibinfo{title}{{Non-hysteretic colossal magnetoelectricity in a
  collinear antiferromagnet}}.
\newblock \emph{\bibinfo{journal}{Nat. Commun.}} \textbf{\bibinfo{volume}{5}},
  \bibinfo{pages}{3201} (\bibinfo{year}{2014}).

\bibitem{Kim2015}
\bibinfo{author}{Kim, J.~W.} \emph{et~al.}
\newblock \bibinfo{title}{{Successive magnetic-field-induced transitions and
  colossal magnetoelectric effect in ${\mathrm{Ni}}_{3}{\mathrm{TeO}}_{6}$}}.
\newblock \emph{\bibinfo{journal}{Phys. Rev. Lett.}}
  \textbf{\bibinfo{volume}{115}}, \bibinfo{pages}{137201}
  (\bibinfo{year}{2015}).

\bibitem{Yokosuk2015}
\bibinfo{author}{Yokosuk, M.~O.} \emph{et~al.}
\newblock \bibinfo{title}{{Tracking the continuous spin-flop transition in
  ${\mathrm{Ni}}_{3}{\mathrm{TeO}}_{6}$ by infrared spectroscopy}}.
\newblock \emph{\bibinfo{journal}{Phys. Rev. B}} \textbf{\bibinfo{volume}{92}},
  \bibinfo{pages}{144305} (\bibinfo{year}{2015}).

\bibitem{Lass2020}
\bibinfo{author}{Lass, J.} \emph{et~al.}
\newblock \bibinfo{title}{{Field-induced magnetic incommensurability in
  multiferroic ${\mathrm{Ni}}_{3}{\mathrm{TeO}}_{6}$}}.
\newblock \emph{\bibinfo{journal}{Phys. Rev. B}}
  \textbf{\bibinfo{volume}{101}}, \bibinfo{pages}{054415}
  (\bibinfo{year}{2020}).

\bibitem{Yokosuk2016}
\bibinfo{author}{Yokosuk, M.~O.} \emph{et~al.}
\newblock \bibinfo{title}{{Magnetoelectric coupling through the spin flop
  transition in ${\mathrm{Ni}}_{3}{\mathrm{TeO}}_{6}$}}.
\newblock \emph{\bibinfo{journal}{Phys. Rev. Lett.}}
  \textbf{\bibinfo{volume}{117}}, \bibinfo{pages}{147402}
  (\bibinfo{year}{2016}).

\bibitem{Zhou2021}
\bibinfo{author}{Zhou, X.}, \bibinfo{author}{Feng, W.}, \bibinfo{author}{Yang,
  X.}, \bibinfo{author}{Guo, G.-Y.} \& \bibinfo{author}{Yao, Y.}
\newblock \bibinfo{title}{Crystal chirality magneto-optical effects in
  collinear antiferromagnets}.
\newblock \emph{\bibinfo{journal}{Phys. Rev. B}}
  \textbf{\bibinfo{volume}{104}}, \bibinfo{pages}{024401}
  (\bibinfo{year}{2021}).

\bibitem{Sell1967}
\bibinfo{author}{Sell, D.~D.}, \bibinfo{author}{Greene, R.~L.} \&
  \bibinfo{author}{White, R.~M.}
\newblock \bibinfo{title}{{Optical exciton-magnon absorption in
  Mn${\mathrm{F}}_{2}$}}.
\newblock \emph{\bibinfo{journal}{Phys. Rev.}} \textbf{\bibinfo{volume}{158}},
  \bibinfo{pages}{489--510} (\bibinfo{year}{1967}).

\bibitem{Greene1965}
\bibinfo{author}{Greene, R.~L.}, \bibinfo{author}{Sell, D.~D.},
  \bibinfo{author}{Yen, W.~M.}, \bibinfo{author}{Schawlow, A.~L.} \&
  \bibinfo{author}{White, R.~M.}
\newblock \bibinfo{title}{{Observation of a spin-wave sideband in the optical
  spectrum of ${\mathrm{MnF}}_{2}$}}.
\newblock \emph{\bibinfo{journal}{Phys. Rev. Lett.}}
  \textbf{\bibinfo{volume}{15}}, \bibinfo{pages}{656--659}
  (\bibinfo{year}{1965}).

\bibitem{Chen2011}
\bibinfo{author}{Chen, P.}, \bibinfo{author}{Lee, N.}, \bibinfo{author}{McGill,
  S.}, \bibinfo{author}{Cheong, S.-W.} \& \bibinfo{author}{Musfeldt, J.~L.}
\newblock \bibinfo{title}{{Magnetic-field-induced color change in
  $\ensuremath{\alpha}$-Fe${}_{2}$O${}_{3}$ single crystals}}.
\newblock \emph{\bibinfo{journal}{Phys. Rev. B}} \textbf{\bibinfo{volume}{85}},
  \bibinfo{pages}{174413} (\bibinfo{year}{2012}).

\bibitem{Henderson1989}
\bibinfo{author}{Henderson, B.} \& \bibinfo{author}{Imbusch, G.~F.}
\newblock \emph{\bibinfo{title}{{Optical spectroscopy of inorganic solids}}}
  (\bibinfo{publisher}{Oxford [Oxfordshire]; Clarendon Press},
  \bibinfo{year}{1989}).

\bibitem{Huang1950}
\bibinfo{author}{Huang, K.} \& \bibinfo{author}{Rhys, A.}
\newblock \bibinfo{title}{{Theory of light absorption and non-radiative
  transitions in $F$-centres}}.
\newblock \emph{\bibinfo{journal}{Proceedings of the Royal Society of London.
  Series A, Mathematical and Physical Sciences}}
  \textbf{\bibinfo{volume}{204}}, \bibinfo{pages}{406--423}
  (\bibinfo{year}{1950}).

\bibitem{Liu2020}
\bibinfo{author}{Liu, Z.} \emph{et~al.}
\newblock \bibinfo{title}{{Observation of nonreciprocal magnetophonon effect in
  nonencapsulated few-layered CrI$_3$}}.
\newblock \emph{\bibinfo{journal}{Sci. Adv.}} \textbf{\bibinfo{volume}{6}},
  \bibinfo{pages}{eabc7628} (\bibinfo{year}{2020}).

\bibitem{Nomura2019}
\bibinfo{author}{Nomura, T.} \emph{et~al.}
\newblock \bibinfo{title}{Phonon magnetochiral effect}.
\newblock \emph{\bibinfo{journal}{Phys. Rev. Lett.}}
  \textbf{\bibinfo{volume}{122}}, \bibinfo{pages}{145901}
  (\bibinfo{year}{2019}).

\bibitem{Leung1991}
\bibinfo{author}{Leung, F.}, \bibinfo{author}{Chiu, W.} \&
  \bibinfo{author}{Demokan, M.}
\newblock \bibinfo{title}{Fiber-optic current sensor developed for power system
  measurement}.
\newblock In \emph{\bibinfo{booktitle}{1991 International Conference on
  Advances in Power System Control, Operation and Management, APSCOM-91.}},
  \bibinfo{pages}{637--643 vol.2} (\bibinfo{year}{1991}).

\bibitem{Guan2010}
\bibinfo{author}{{Guan}, B.} \& \bibinfo{author}{{Wang}, S.}
\newblock \bibinfo{title}{Fiber grating laser current sensor based on magnetic
  force}.
\newblock \emph{\bibinfo{journal}{IEEE Photon. Technol. Lett.}}
  \textbf{\bibinfo{volume}{22}}, \bibinfo{pages}{230--232}
  (\bibinfo{year}{2010}).

\bibitem{Wen2012}
\bibinfo{author}{Wen, F.}, \bibinfo{author}{Wu, B.-J.}, \bibinfo{author}{Li,
  C.-Z.}, \bibinfo{author}{Wu, S.-J.} \& \bibinfo{author}{Perumal, S.}
\newblock \bibinfo{title}{{Magnetic field response of erbium-doped
  magneto-optic fiber Bragg grating}}.
\newblock \emph{\bibinfo{journal}{Opt. Eng.}} \textbf{\bibinfo{volume}{51}},
  \bibinfo{pages}{1 -- 5} (\bibinfo{year}{2012}).

\bibitem{Nascimento2015}
\bibinfo{author}{Nascimento, I.~M.}, \bibinfo{author}{Baptista, J.~M.},
  \bibinfo{author}{Jorge, P. A.~S.}, \bibinfo{author}{Cruz, J.~L.} \&
  \bibinfo{author}{Andrés, M.~V.}
\newblock \bibinfo{title}{{Erbium doped optical fiber lasers for magnetic field
  sensing}}.
\newblock In \bibinfo{editor}{Kalinowski, H.~J.}, \bibinfo{editor}{Fabris,
  J.~L.} \& \bibinfo{editor}{Bock, W.~J.} (eds.) \emph{\bibinfo{booktitle}{24th
  International Conference on Optical Fibre Sensors}}, vol.
  \bibinfo{volume}{9634}, \bibinfo{pages}{765 -- 768}.
  \bibinfo{organization}{International Society for Optics and Photonics}
  (\bibinfo{publisher}{SPIE}, \bibinfo{year}{2015}).

\bibitem{Zhang2017}
\bibinfo{author}{Zhang, T.}, \bibinfo{author}{Zhang, J.},
  \bibinfo{author}{Cheng, L.}, \bibinfo{author}{Li, Y.} \&
  \bibinfo{author}{Guan, B.-O.}
\newblock \bibinfo{title}{Magnetic field sensing through magnetic force using
  erbium-doped fiber laser}.
\newblock In \emph{\bibinfo{booktitle}{2017 Conference on Lasers and
  Electro-Optics Pacific Rim}}, \bibinfo{pages}{s1169}
  (\bibinfo{publisher}{Optical Society of America}, \bibinfo{year}{2017}).

\bibitem{VASP1}
\bibinfo{author}{Kresse, G.} \& \bibinfo{author}{Hafner, J.}
\newblock \bibinfo{title}{{\textit{Ab initio} molecular dynamics for liquid
  metals}}.
\newblock \emph{\bibinfo{journal}{Phys. Rev. B}} \textbf{\bibinfo{volume}{47}},
  \bibinfo{pages}{558--561} (\bibinfo{year}{1993}).

\bibitem{VASP2}
\bibinfo{author}{Kresse, G.} \& \bibinfo{author}{Furthm\"uller, J.}
\newblock \bibinfo{title}{Efficient iterative schemes for \textit{ab initio}
  total-energy calculations using a plane-wave basis set}.
\newblock \emph{\bibinfo{journal}{Phys. Rev. B}} \textbf{\bibinfo{volume}{54}},
  \bibinfo{pages}{11169--11186} (\bibinfo{year}{1996}).

\bibitem{Dudarev1998}
\bibinfo{author}{Dudarev, S.~L.}, \bibinfo{author}{Botton, G.~A.},
  \bibinfo{author}{Savrasov, S.~Y.}, \bibinfo{author}{Humphreys, C.~J.} \&
  \bibinfo{author}{Sutton, A.~P.}
\newblock \bibinfo{title}{{Electron-energy-loss spectra and the structural
  stability of nickel oxide: An LSDA+U study}}.
\newblock \emph{\bibinfo{journal}{Phys. Rev. B}} \textbf{\bibinfo{volume}{57}},
  \bibinfo{pages}{1505--1509} (\bibinfo{year}{1998}).

\bibitem{PBEsol}
\bibinfo{author}{Csonka, G.~I.} \emph{et~al.}
\newblock \bibinfo{title}{Assessing the performance of recent density
  functionals for bulk solids}.
\newblock \emph{\bibinfo{journal}{Phys. Rev. B}} \textbf{\bibinfo{volume}{79}},
  \bibinfo{pages}{155107} (\bibinfo{year}{2009}).

\bibitem{wien2k}
\bibinfo{author}{Blaha, P.}, \bibinfo{author}{Schwarz, K.},
  \bibinfo{author}{Madsen, G. K.~H.}, \bibinfo{author}{Kvasnicka, D.} \&
  \bibinfo{author}{Luitz, J.}
\newblock \emph{\bibinfo{title}{WIEN2k, an augmented plane wave + local
  orbitals program for calculating crystal properties}}
  (\bibinfo{publisher}{Karlheinz Schwarz, Techn. Universit\"{a}t Wien,
  Austria}, \bibinfo{year}{2001}).

\bibitem{eDMFT}
\bibinfo{author}{Haule, K.}
\newblock \bibinfo{title}{Structural predictions for correlated electron
  materials using the functional dynamical mean field theory approach}.
\newblock \emph{\bibinfo{journal}{J. Phys. Soc. Jpn}}
  \textbf{\bibinfo{volume}{87}}, \bibinfo{pages}{041005}
  (\bibinfo{year}{2018}).

\end{thebibliography}


\section*{Figure Legends}

\end{document}